# Energetic Electron Transport in Magnetic Fields with Island Chains and Stochastic Regions

E G Kostadinova [1], D M Orlov [2], M Koepke [3], F Skiff [4], M E Austin [5]

[1] *Physics Department, Auburn University, AL, USA*

[2] *Center for Energy Research, UC San Diego, La Jolla, CA, USA*

[3] *Department of Physics and Astronomy, West Virginia University, Morgantown, WV*

[4] *Department of Physics and Astronomy, University of Iowa, Iowa City, IA*

[5] *Institute for Fusion Studies, The University of Texas at Austin, Austin, TX*

Contact e-mail: egk0033@auburn.edu

**Abstract.** This paper investigates energetic electron transport in magnetized toroidal plasmas with magnetic fields characterized by island chains and regions of stochastic field lines produced by coil perturbations. We report on experiments performed in the DIII-D tokamak, which utilize electron cyclotron heating and current drive pulses to 'tag' electron populations within different locations across the discharge. The cross-field transport of these populations is then inferred from electron cyclotron emission measurements and gamma emission signals from scintillator detectors. Two types of energetic particles are distinguished and discussed: nonrelativistic suprathermal electrons and relativistic runaway electrons. The magnetic field topology in each discharge is reconstructed with field line tracing codes, which are also used to determine the location and scale of magnetic islands and stochastic regions. Comparison of simulations and experiments suggests that suprathermal transport is suppressed when the tagging is performed at a smaller radial location than the location of the $q = 1$ island chain, and enhanced otherwise. We further demonstrate that increasing the width of the stochastic region within the edge plasma yields enhancement of the suprathermal electron transport.

**Keywords:** anomalous diffusion, energetic electrons, DIII-D tokamak, magnetic islands, field stochasticity

# I. INTRODUCTION

An island in a magnetic field topology is a closed magnetic flux tube, characterized by a central elliptic point, called an O-point, hyperbolic X-points, and a separatrix surface, isolating the structure from the rest of space (Fig. 1a.) Particles within the separatrix circulate around the elliptic point in a bound trajectory with a resonant frequency proportional to the island width, while particles outside the separatrix are not resonant but still experience periodic changes in energy and momentum [1]. Regions of stochastic magnetic field lines can occur when homoclinic tangles form near the island X-points [2], or when neighboring island chains overlap [3]–[5], or when a single island chain bifurcates into sub-island chains [6]–[8]. Stochasticity of the magnetic field lines can enhance cross-field transport due to chaotic particle trajectories [9]. In all three stochasticity mechanisms mentioned here (X-point-proximity tangles, island overlap, and island bifurcation), the stochastic region coexists with remnants of magnetic islands (Fig. 1b), that can act as attractors [10]. In these circumstances, nonlocal interactions occur simultaneously with chaotic trajectories, and the resulting particle transport can be subdiffusive, diffusive, or superdiffusive [11]. The plots in Fig. 1 were generated using the TRIP3D code [12], [13].

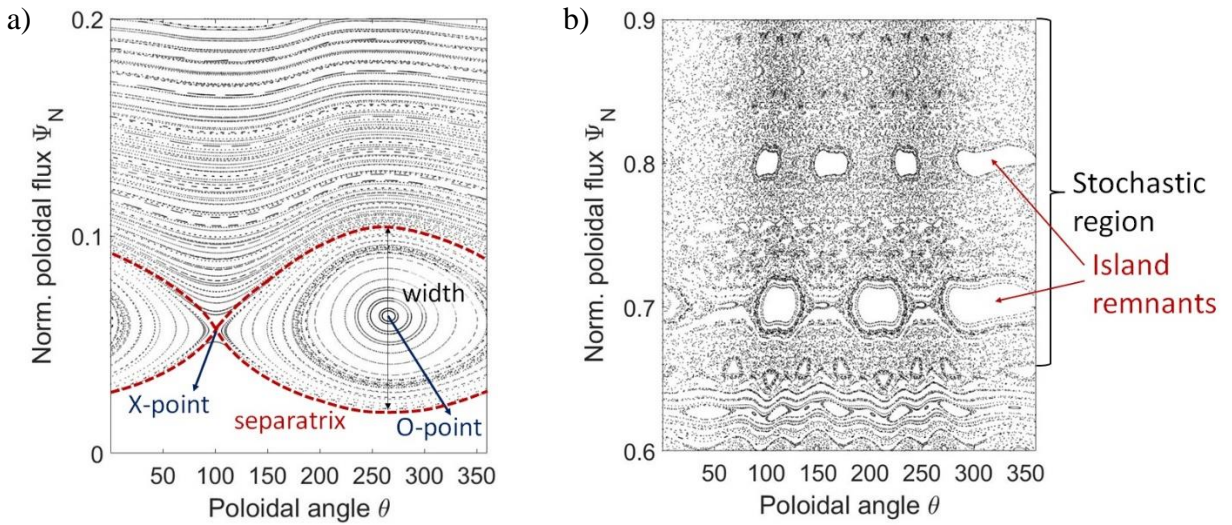

Fig. 1. Poincaré plots of DIII-D shot #172330: a) The structure of a magnetic island chain in the core plasma. The separatrix is drawn by a red dashed line to guide the eye. b) Edge plasma stochastic region along with remnants of magnetic islands during island overlap induced by coil perturbations.

Magnetic-island structures and field-line stochasticity are ubiquitous in the inhomogeneous and nonstationary space environment (e.g., solar wind plasma [14], [15] and Earth's magnetosphere [16]–[18]) and in controlled laboratory settings (e.g., tokamaks [19], [20] and stellarators [21], [22]). Since both features can originate from the dynamical processes of magnetic reconnection and turbulence, their concurrence is common in magnetized plasma [23]. An outstanding research question in space plasma and magnetically confined fusion is understanding the cause and effect of such complexity in the magnetic field topology. Specifically, it is not clear how the presence of either island chains or stochasticity influences particle acceleration to suprathermal energies, and whether the interplay between the two features enhances or diminishes cross-field transport for a given scenario. An example approach to studying these questions is considering a toroidal magnetized plasma experiment where the number, size, location, and overlap of magnetic islands can be controlled using small-amplitude Resonant Magnetic Perturbations (RMPs).

The effects of magnetic field perturbations on particle transport have been previously investigated in tokamak experiments where RMPs were used to control plasma-wall interactions and edge localized modes (ELMs) [3], [24]–[26]. However, these studies have typically focused on high confinement (H-mode) plasmas, where the plasma response to the external perturbation fields is large – complicating the analysis. As in many contexts in plasma physics where low-dimensional chaos may occur, little is known about the complex dynamics resulting from the self-consistent back reaction of collective effects on the field-line structure and the chaos-induced transport which presumably causes the plasma response. This necessitates the development of comprehensive three-dimensional simulations, capturing nonlinear plasma response. Yet, most simulations of the [27]MHD modeling using a single-toroidal-helicity applied magnetic perturbation [28], [29]. It has been identified that the development of advanced simulations and the interpretation of H-mode experimental results can benefit from improved analytical models informed from low-confinement (L-mode) plasma experiments where the plasma response is minimized [30].

This paper presents an experimental study of non-thermal electron transport in L-mode toroidal plasma with magnetic islands and regions of stochastic magnetic field lines. Here we analyze DIII-D experiments conducted in the DIII-D tokamak [30], where 3D non-axisymmetric RMPs were created using two rows of six window-frame coils located off-midplane inside the vessel [31]. These coils were used to create and grow magnetic islands at the $n = 3$ rational surfaces (i.e., at safety factor ratios $q = m/n$, where $n = 3$) and to cause the controlled formation of stochastic regions in the edge plasma through island chain overlap. These experiments featured low density, inner-wall limited L-mode discharges with no neutral beam injection, using only Ohmic heating. In such conditions, the nonlinear plasma response is expected to be negligible, which allows for the study of electron transport as a function of the magnetic field topology constructed from vacuum field simulations. Here we consider small-amplitude magnetic field perturbations $\delta B$, such that $(\delta B)/B_T \lesssim 10^{-4}$, where $B_T$ is the toroidal magnetic field. A main finding of this study is that suprathermal electron transport is enhanced or suppressed depending on electron location with respect to the large magnetic islands in the core plasma. We also established that the prominence of the suprathermal electron feature is proportional to the width of the stochastic region in the edge plasma.

For each examined $B$-field topology, to infer the cross-field electron transport, we implemented an electron 'tagging' technique, which has been previously used to study ion acceleration in linear plasma [32] and suprathermal electrons in toroidal plasma [33]. Electron Cyclotron Heating (ECH) and Current Drive (ECCD) pulses were employed to produce a temporally modulated and spatially localized enhanced current leading to a perturbation of the local electron population in a desired resonance layer within the plasma. Then, the time-dependent radial spreading of the 'tagged' electrons was observed from Electron Cyclotron Emission (ECE) measurements. Normally, it is expected that due to resonance with the ECH wave frequency, the tagged electrons will heat up and then thermalize, leading to a uniform increase in the plasma temperature profile. However, if the resonant layer contains suprathermal electrons, the tagging causes them to escape from the plasma before thermalization can occur. The latter results in ECE profiles where the edge plasma channels measure considerably higher electron temperature than what is calculated from Thomson Scattering (TS) data. Since TS fits assume Maxwellian (thermal) electron distributions, the observed deviations are interpreted as signatures of nonthermal electrons, as discussed in Sec. II.

In addition to ECE profiles, gamma emission signals from scintillator detectors were used to determine if the observed non-thermal electrons were also relativistic. Runaway electrons, traveling at relativistic speeds, were detected during the first ECH/ECCD pulse in several discharges, but not during subsequent ECH/ECCD pulses. A possible explanation is that runaways are generated and confined in the core plasma during current ramp up at the beginning of each discharge, and then de-confined by the interaction with the first ECH/ECCD pulse. Interestingly, in each discharge, suprathermal features in the ECE measurements were observed during all ECH/ECCD pulses, but no peaks in the scintillator data were recorded after the first pulse, indicating that electrons detected during subsequent pulses were non-thermal but also non-relativistic. Thus, in this study, we could distinguish between two types of energetic particles: nonrelativistic suprathermal electrons and runaway electrons.

This paper is organized as follows. Section II introduces the details of the experimental setup and presents the main results; namely, the observation of energetic electrons from several diagnostic measurements. In Sec. III, the magnetic field topology corresponding to the experimental conditions is obtained from simulations of the vacuum magnetic field. These simulations are used to identify the number, size, and location of magnetic islands. In this section, we also discuss the 3D coil current threshold needed to induce island overlap in the edge plasma, resulting in a region of stochastic magnetic field lines. Section IV provides interpretation of the results from the comparison between simulations and experiments. Specifically, we discuss the possible mechanisms leading to the presence of energetic electrons for each magnetic field topology. In Sec. V, we summarize the conclusions and discuss future work.

## II. EXPERIMENTAL RESULTS

### A. Electron 'Tagging' Experiments on DIII-D

Here we consider ten discharges, shot numbers 172321-172330, which were part of the 2017 Frontiers in Science campaign at DIII-D. Each shot was an inner-wall limited L-mode discharge (Fig 2a) with no neutral beam injection, using only Ohmic heating and modulated ECH and ECCD pulses during the plasma current $I_p$ plateau (Fig. 2b). In each shot, the ECCD is obtained by toroidally aiming the ECH launchers (as opposed to only radially aimed, which provides heating only, and no ECCD.) Typical parameters for this campaign were $I_p = (0.7 - 1.1)\ MA$, toroidal magnetic field $B_t \approx 2\ T$, and edge safety factor $q_{95} = (4.8 - 7.4)$. Note that two of the current profiles in Fig. 2b, shot 172325 (orange line) and shot 172326 (light green line) drop sharply at time $\gtrsim 5000\ ms$ due to plasma disruption. Thus, the present analysis only includes data for time $< 5000\ ms$. The $q_{95}$ factor provides the ratio between the poloidal winding number $m$ and the toroidal winding number $n$ of the magnetic field lines on the surface corresponding to the 95% of the magnetic flux. For these discharges, the ratio $m/n$ increases monotonically from the core to the edge surfaces. The $q_{95}$ factor is a useful measure quantifying the number of rational surfaces, and hence the number of magnetic islands, that can be supported in the edge plasma. Higher $q_{95}$ corresponds to increased number of closely spaced surfaces and higher $q_{95}$ shear in the edge, which in turn yields higher number of islands and increased probability for overlap of neighboring island chains. However, lower $q_{95}$ can support wider islands in the edge plasma, which can result in wider stochastic edge region when RMP is applied. This will be further discussed in Sec III B.

To minimize plasma response, measures were taken to maintain low plasma densities, $n_e = (0.9 - 1.6) \times 10^{19} \, m^{-3}$, without triggering locked modes (standard $n = 1$ Error Field Correction applied via the ex-vessel C-coils), or sawtooth instabilities after $t = 1000 \, ms$. For the simple inner-wall limited geometry, low plasma density, and lack of toroidal plasma rotation (due to no neutral beam heating), the plasma response is likely weak, which allows to infer the relevant transport mechanisms from simulations of the vacuum magnetic field topology (as discussed in Sec. III A.) In each shot, a magnetic field perturbation was induced using DIII-D in-vessel I-coils [25], [34], whose current amplitude was varied in the range $I_{RMP} = \pm(1 - 6) \, kA$ (see Fig 2c and Table I.) An important goal for these experiments was to measure changes in electron transport as a function of RMP coil current and link them to the amount of magnetic field stochasticity produced by the perturbation. In Sec. III B we will demonstrate that increasing the width of the stochastic edge region yields an increase in the observed suprathermal electron transport.

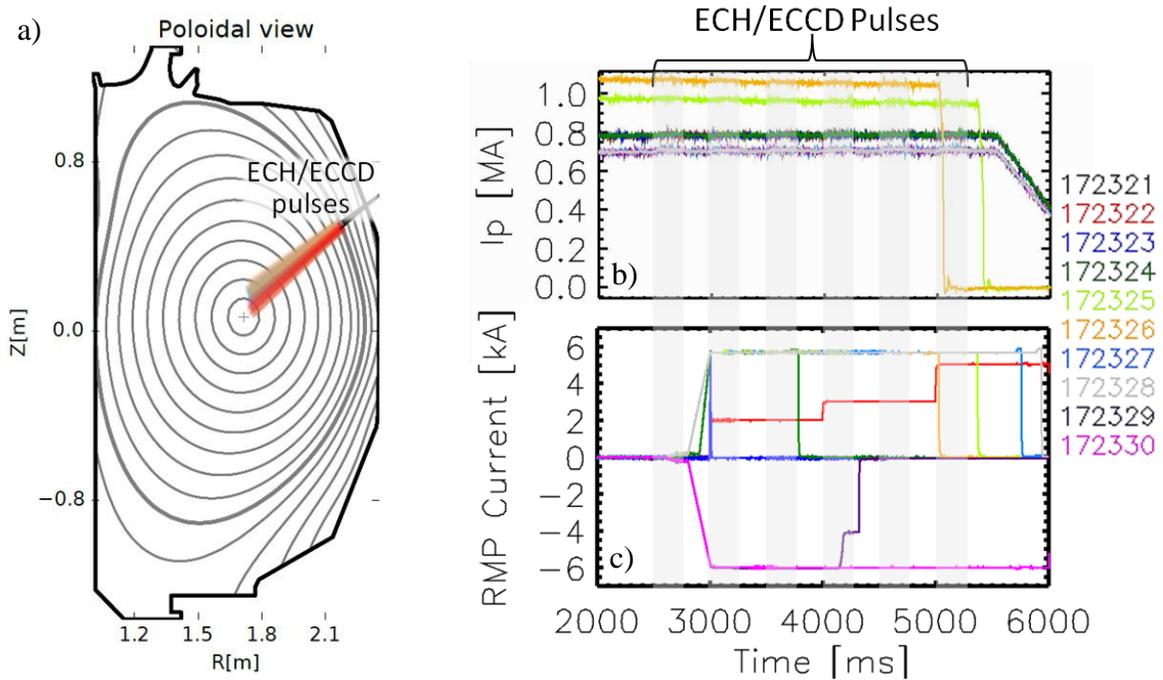

Fig. 2. a) Inner-wall limited L-mode discharge plasma shape for shot 172330 at 2600 ms showing an ECH/ECCD pulse consisting of electromagnetic waves from six gyrotrons (colored lines on the plot). The solid lines on the plot represent magnetic flux surfaces. b) Plasma current $I_p$ and c) RMP coil current $I_{RMP}$ time traces for shots 172321-172330. The grey shaded rectangles in b) and c) mark the time intervals of the ECH/ECCD pulses.

To quantify transport, the experiments utilized a technique called 'electron tagging', in which an ECH/ECCD pulse produces a temporally modulated and spatially localized enhanced current at a specific location within the discharge. In cyclotron heating, resonance occurs between the cyclotron frequency $\omega_c$ of electrons and the frequency $\omega_{ext}$ of external beam of electromagnetic radiation. At DIII-D, the frequency of the gyrotrons is fixed at 110 GHz, which corresponds to a vertical resonant layer extending from the core to the edge plasma. The heating location is adjusted by aiming the gyrotrons so that the ECH/ECCD beam intersects the (~vertical) resonance layer at the desired toroidal flux/radial location, so that electrons at that location receive a 'kick' from the electromagnetic beam. Then, the time-dependent radial spreading of the electron perturbation can be observed from Thomson scattering and Electron Cyclotron Emission (ECE) measurements. Each experimental shot discussed here included six ECH/ECCD pulses with typical pulse duration of $\approx 250\ ms$ and amplitudes in the range $(1.1 - 3.4)\ MW$. The six pulses occurred at fixed times in each discharge (see grey rectangles on Fig 2b, c), which allows for meaningful comparison across discharges. Table I lists parameters relevant to the ECH/ECCD tagging and RMP perturbations used in the experiments. Here $I_{RMP}$ is the $n = 3$ RMP coil current amplitude, $P_{ABS}$ is the integrated power absorbed in each pulse, $I_{ECCD}$ is the integrated current drive in each pulse, and $\rho_{ECCD}$ is the peak (normalized) radial location of the current drive. In all shots, the ECH pulse frequency was $\nu_{ECH} = 110\ GHz$, and the ECH polarization was X-mode second harmonic. The selected time slices feature different combinations of perturbation current, ECH absorbed power, and ECCD current. All values in Table I were obtained using TORAY [35]. TORAY is a ray tracing code for studying electron cyclotron heating and current drive in toroidal geometry. Figure 3 shows typical TORAY plots of the radial distribution of ECH heat flux and ECCD current flux, together with the integrated values used in Table I.

Table I. ECH/ECCD parameters for shots 172322-172330 obtained from TORAY code.

| Shot | $t\ [ms]$ | $I_{RMP}\ [kA]$ | $P_{ABS}\ [MW]$ | $I_{ECCD}\ [kA]$ | $\rho_{ECCD}$ |
|---|---|---|---|---|---|
| **172322** | 2600 | 0 | 2 | -54.4 | 0.225 |
| **172322** | 3100 | 2 | 2 | -45 | 0.224 |
| **172322** | 4100 | 3 | 2.1 | -49.5 | 0.222 |
| **172322** | 5100 | 5 | 2.1 | -48.6 | 0.226 |
| **172323** | 2600 | 0 | 1.5 | 62.6 | 0.205 |
| **172323** | 3100 | 0 | 1.1 | 31.2 | 0.226 |
| **172324** | 2600 | 0 | 2.2 | 105.1 | 0.186 |
| **172324** | 3100 | 5.7 | 1.7 | 57.5 | 0.203 |
| **172325** | 2600 | 0 | 2.2 | 91.4 | 0.216 |
| **172325** | 3100 | 5.7 | 2.3 | 86.3 | 0.224 |
| **172326** | 2600 | 0 | 2.2 | 82.2 | 0.229 |
| **172326** | 3100 | 5.7 | 2.2 | 72.3 | 0.242 |
| **172327** | 2600 | 0 | 2.2 | 181.7 | 0.153 |
| **172327** | 3100 | 5.7 | 2.3 | 168.8 | 0.157 |
| **172328** | 2600 | 0 | 2.9 | 485.6 | 0 |
| **172328** | 3100 | 5.7 | 2.9 | 374.2 | 0 |
| **172329** | 2600 | 0 | 2.8 | 409.1 | 0.005 |
| **172329** | 3100 | -6 | 2.8 | 393.9 | 0 |
| **172330** | 2600 | 0 | 3.4 | 245 | 0.144 |
| **172330** | 3100 | -6 | 2.9 | 192.1 | 0.132 |

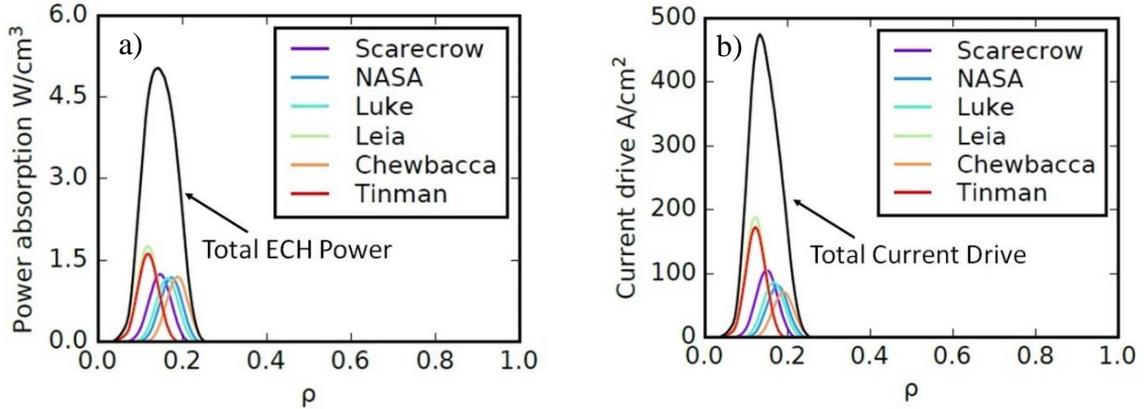

Fig. 3. Shot 172330 at $2600\ ms$: radial distribution of a) ECH power absorption, b) ECCD current drive, obtained using TORAY code. Different colors correspond to the six gyrotrons available at DIII-D. The curve in black in each shot represents the total values.

B. Observation of Suprathermal Electron Transport

The presence of suprathermal electrons can be inferred from an unusual increase in electron temperature in the edge plasma at normalized radius $\rho \gtrsim 1$, observed in ECE radiometer measurements, as shown in Fig. 4. What is observed in the present experiments is a milder case of the more extreme example of non-thermal ECE shown by Harvey et al. [36] (In general, $\rho$ is the square root of the normalized toroidal flux surface.) The DIII-D ECE radiometer [37] is a multichannel heterodyne system that provides electron temperature from measurements of optically thick, second harmonic electron cyclotron emission. The instrument viewing is along a horizontal chord at the tokamak midplane at a toroidal angle of 81 degrees. If the radial electron transport evolves according to classical diffusion, the electron temperature measured by ECE cords is expected to decrease with increasing radial location, following a similar pattern as the Thompson scattering [38] data (dashed green lines on Fig. 4). In all examined shots, the electron temperature measured by ECE cord 01 was consistently higher than the temperature measured by ECE cord 02, even though cord 01 is located at larger radial location $\rho$ than cord 02.

Vertical green lines in Fig. 4 indicate the location of several rational surfaces, while the vertical red line in Fig. 4b indicates the location of the ECH/ECCD tagging for that discharge. The safety factor, or $q$ profile, and the location of the $q = 1$ and $q = 2$ surfaces mentioned throughout the paper are inferred from the magnetics-only Grad-Shafranov equilibrium solution. As no neutral beams were used in these discharges to keep the plasma rotation and plasma response minimized, the motional Stark effect (MSE) measurements of the pitch angle in the plasma core were not available for the shots presented here.

Figure 4 shows a typical ECE profile from shot 172330 before and during the ECH/ECCD pulse. While in both cases, the temperature measured by cord 01 is higher than the one measured by cord 02, the tagging technique clearly enhances the observed suprathermal ECE tail. Comparison between Fig. 4a and Fig. 4b reveals that during the tagging, the entire ECE radial profile shifts to higher temperatures (note the difference in the abscissa) but the tail feature and the flat region around the $q = 2$ surface are preserved. The flattening of the ECE profile around the $q = 2$ surface, which is observed for all examined discharges, can be associated with a 2/1 island located on that surface. A main difference between Fig. 4a and Fig. 4b is that prior to the pulse, the temperature measurements around the $q = 1$ surface are somewhat flat, while during the pulse, the temperature profile in the same region exhibits a steep slope. Flat regions around the $q = 1$ surface can be associated with the presence of a sawtooth instability or of another magnetic island chain. Since the conditions in the present discharges were carefully selected to minimize plasma response, a sawtooth instability was not observed for these shots, suggesting that the flat region is caused by a $q = 1$ island chain.

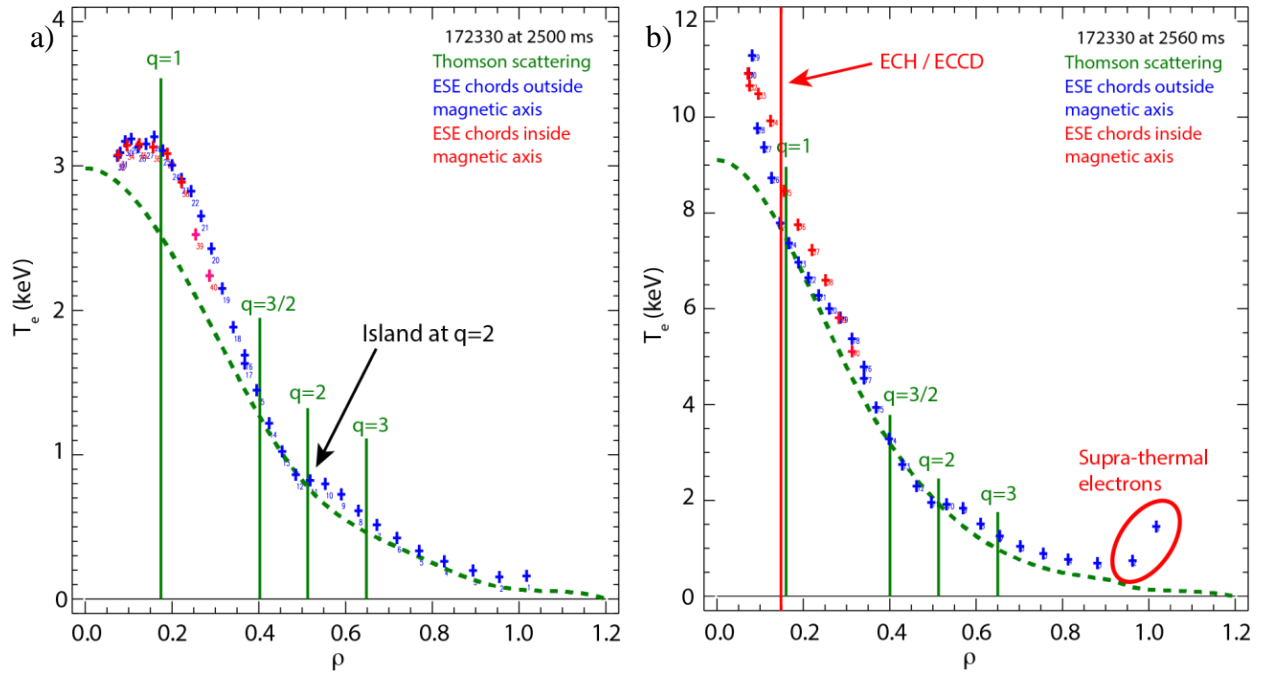

Fig. 4. Radial distribution of electron temperature from ECE for shot 172330 a) at $2500\ ms$ prior to an ECH/ECCD pulse b) at $2560\ ms$ during an ECH/ECCD pulse. Dashed green lines represent electron temperature from a fit to the Thomson scattering data, which assume thermal electron distribution function.

A likely interpretation of the observed changes around the $q = 1$ surface is that when the ECH/ECCD pulse is delivered within the island chain located there, the pulse shrinks the island by breaking up the island separatrix, which, in turn, leads to the enhanced release of energetic electrons [39]–[41]. Examination of similar ECE distribution plots for the other discharges leads to several observations: (i) the change around the $q = 1$ surface during the tagging is sensitive to ECH power, (ii) the first ECH/ECCD pulse in each shot leads to the most prominent suprathermal tail on the ECE distribution, and (iii) the suprathermal tail is present with and without the RMP coil perturbations. However, in Sec. IV, we discuss how the RMP coil perturbations indirectly affect the observed electron transport through the formation of stochastic edge plasma region.

Figure 5 shows a plot of the highest ECE temperature measured in the core plasma as a function of ECH power during the first ECH pulse for each shot. Although the deposition location and current drive slightly vary across shots (see Table I), a clear trend is observed showing that electron peak temperature scales almost linearly with ECH power. Note that shots 172325 (light green) and 172326 (orange) have the lowest $q_{95}$ values of $\approx 5.2$ and $\approx 4.7$, respectively, while shots 172323 (light blue) and 172328 (grey) have the highest $q_{95}$ of $\approx 7.3$. Thus, we also expect that for a fixed ECH power and similar deposition locations (see Table I), the maximum electron temperature will increase with increasing $q_{95}$, likely due to decreased density (as further discussed in Sec. IV.A.)

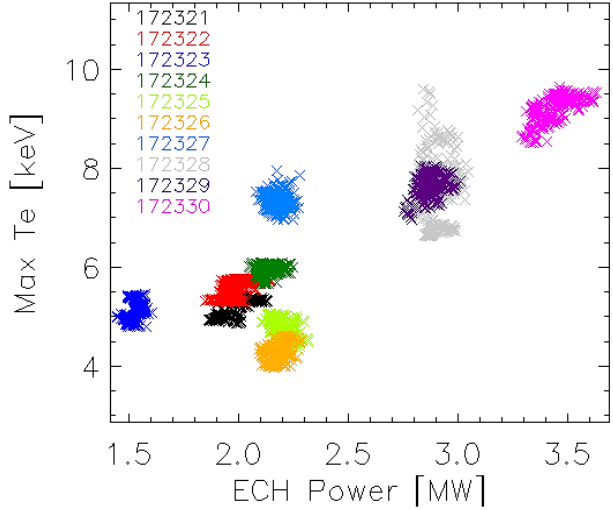

Fig. 5. Maximum electron temperature (from ECE chord located in the core plasma) versus ECH power for period $2550\ ms - 2600\ ms$.

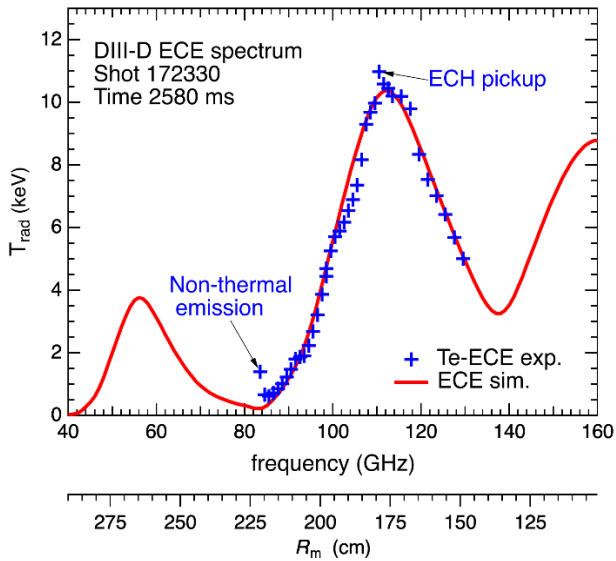

Fig. 6. Electron temperature as a function of frequency for shot 172330 at $2580\ ms$. Solid red line shows the predicted values from ECE simulations assuming thermal electron distribution. Blue crosses mark experimental data from ECE measurements. The second axis under the plot shows the location of the ECE chords as a function of major axis $R_m$.

To verify that the observed increase in ECE measurements at $\rho \gtrsim 1$ is not caused by the drop of plasma density outside the separatrix, ECE simulations were conducted (using the ECESIM code [42]) to obtain the expected electron temperature as a function of frequency expected for a Maxwellian electron distribution. Figure 6 shows the simulation results for discharge 172330 at $2580\ ms$ (solid red line), together with the ECE measured values (blue crosses). The ECE simulations were informed by Thomson scattering data. Deviation from predicted ECE temperature is clearly visible for frequencies $< 84$ GHz. With the low-density L-mode edge, the ECE is marginally blackbody from the region inside the LCFS (last closed flux surface), where channels $>1$ reside, and optically thin for the emission measured by channel 1. Nevertheless, as the simulation shows, one does not expect the edge emission to be elevated if the electron distribution is Maxwellian. Thus, this high temperature measured by ECE is likely caused by suprathermal electrons in the plasma, the radiation escaping due to the low optical thickness of the edge channels.

## C. Runaway Electrons

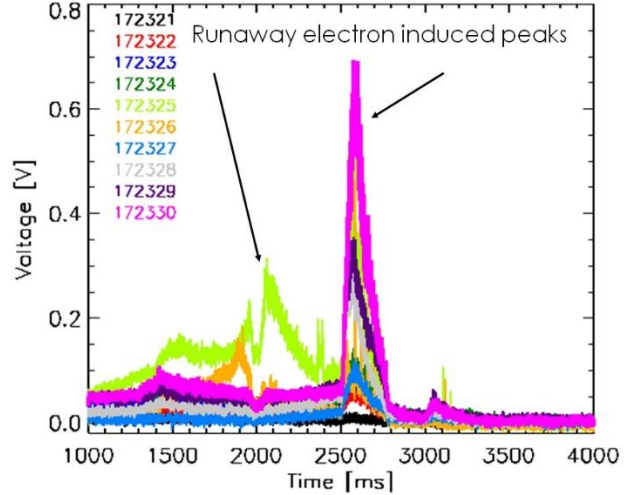

Fig. 7. Runaway-electron-induced gamma emission on the Fast Neutron Scintillator Counter for several shots from the same campaign.

Runaway electrons could be distinguished from the general nonthermal electron population as they also induced gamma emission on fast neutron scintillator counters (Fig. 7). The hard x-ray signals are measured by an uncollimated plastic scintillator placed near the vacuum vessel. Signals from an adjacent ZnS scintillator that is an order of magnitude less sensitive to x-rays [43] confirms that the plastic scintillator is measuring hard x-rays. Interestingly, in each discharge, ECE deviations from the TS fits were observed during all ECH/ECCD pulses, but no peaks in the scintillator data were observed after the first pulse, indicating that electrons detected during subsequent pulses were non-thermal but also not relativistic. The ECE radiometers at DIII-D are preferentially sensitive to non-thermal synchrotron emission from energetic electrons. Due to the radial viewing geometry of the ECE radiometers on DIII-D, these diagnostics probe the high pitch-angle ($60°-90°$) and low energy ($1\ MeV - 5\ MeV$) energetic electron population [44]. The HXR detectors used here cover all pitch angles ($0°-90°$) and electron energy in the range ($5\ MeV - 30\ MeV$). Thus, we expect that the signals showed in Fig. 7 correspond to energetic electrons with energy $\gtrsim 5 MeV$, while the nonthermal electron populations detected by ECE after the first ECH/ECCD pulse have energies in the range ($1\ MeV - 5\ MeV$). Accurate determination of energy and pitch angle is not available in the present experiments, but will be explored in an upcoming experimental campaign, which will utilize multiple additional diagnostics available at DIII-D, including the gamma ray imager [45] and visible synchrotron emission from a tangential camera [46].

The observation of scintillator peaks only during the first ECH/ECCD pulse in some of the examined discharges suggests that the tagging technique can deplete the relativistic runaway population. While this effect has been observed empirically before, the exact mechanism leading to the release of the runaway electrons from the core to the edge plasma is not well understood. Since runaway electrons are accelerated to relativistic speeds likely during the current ramp-up at the beginning of the discharges, we do not expect that their initial transport is strongly affected by the island chain structure or stochastic regions in the magnetic field. However, their transition from strongly confined to lost orbits during an ECH/ECCD pulse can result from spontaneous or induced changes in the magnetic field topology, including stochastization of magnetic field lines near island X-points or island bifurcation. Interestingly, Fig. 7 shows smaller but distinct peaks in the period $1900 - 2300\ ms$ for discharges 172325 and 172326 (light green and orange dots, respectively.) Since there is no sawtooth instability, ECH/ECCD pulse, or 3D coil perturbation in this period, the observed peaks in the scintillator data suggest a spontaneous deconfinement of suprathermal electrons. Spontaneous island bifurcation has been recently observed in DIII-D H-mode experiments for rotating islands in the core plasma [7]. The possibility of similar processes in L-mode discharges and their effect on suprathermal electron transport are topics of interest to a future study.

### D. Suppression of the Suprathermal Feature

In the present experiments, the observed suprathermal features did not appear to change appreciably as the amplitude of the RMP current was varied. For example, at 2600 ms the RMP current for shot 172330 (Fig. 4b) was $I_{RMP} = 0\ kA$; yet the ECE plot exhibits a pronounced suprathermal tail. We note that for all examined cases, the first ECH deposition was performed with no RMP perturbation, and this first pulse always resulted in the largest electron temperature peak and the largest enhancement in the difference between trace 01 and trace 02 measurements. As mentioned in Sec. II C, this effect is likely due to the release of runaway electrons. Thus, in Sec. IV, where we discuss how RMP-induced edge stochasticity affects non-thermal electron transport, we do not consider data from these first ECH/ECCD pulses. In Sec. IV, we show that for fixed properties of the ECH/ECCD pulses and fixed RMP current values, the non-thermal transport depends on the size, location, and distribution of magnetic islands across the plasma.

To demonstrate that the suprathermal electron tail is persistent even after the first pulse (which depletes the runaway electrons), Figure 8 shows the ratio between temperature measured by ECE cords 01 and 02 as a function of time for all examined shots. One expects that $T_{01}/T_{02}$ would be below one for discharges with electron temperature profiles that decrease monotonically in the plasma edge. In the examined discharges, this ratio is bigger than 1 for all cases, except shots 172328 and 172329, where the ECE measurement did not extend beyond $\rho \approx 0.8$ and the outermost ECE channels 02 and 01 were located not in the edge, but in the core of the plasma. Therefore, we omit data from 172328 and 172329 in Fig. 8. While for all other shots, the ratio $T_{01}/T_{02}$ increases during ECH/ECCD pulses (shaded areas in Fig. 8b, c), the effect is reversed for shot 172326 (orange line), as can be seen on Fig. 8c. Shot 172325 (light green line) exhibits transitional behavior, where the ratio is only slightly enhanced at the onset of an ECH/ECCD pulse and gradually decreases over a time period longer than the pulse.

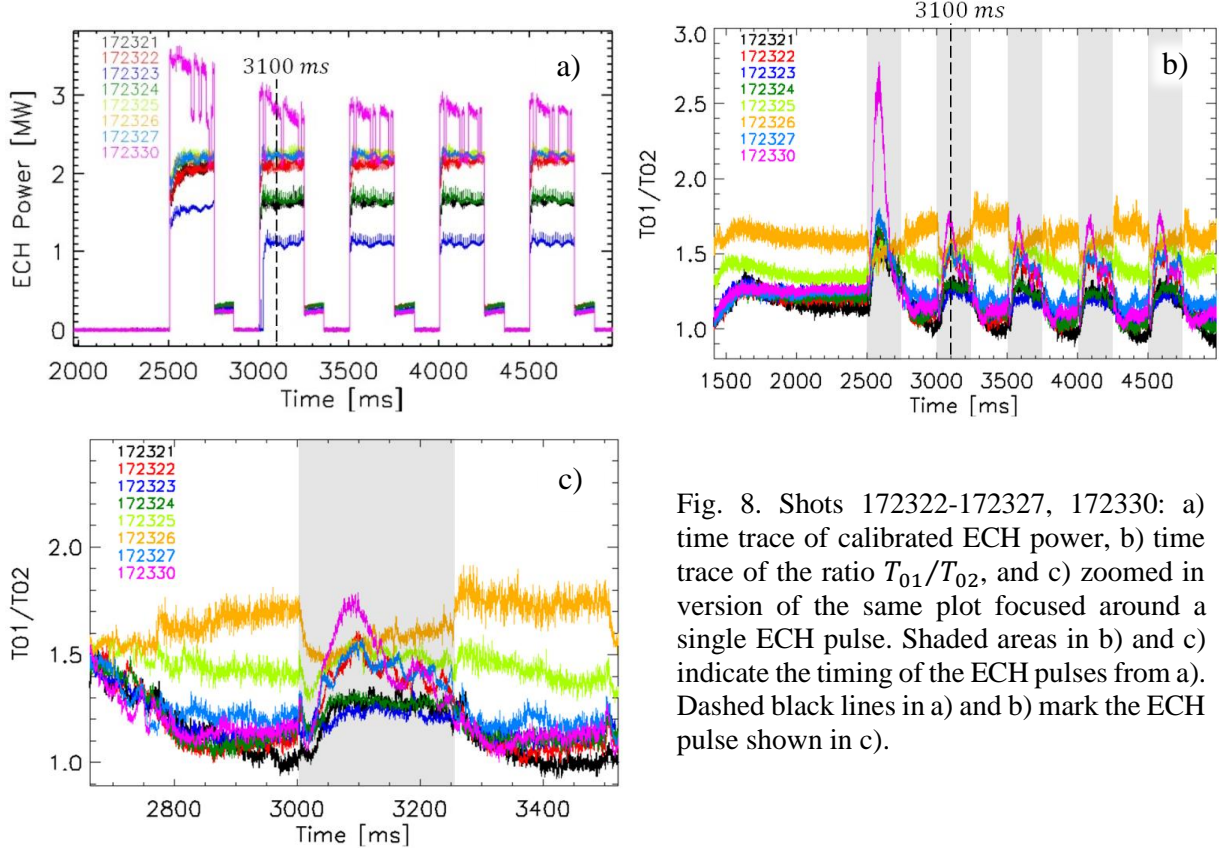

Fig. 8. Shots 172322-172327, 172330: a) time trace of calibrated ECH power, b) time trace of the ratio $T_{01}/T_{02}$, and c) zoomed in version of the same plot focused around a single ECH pulse. Shaded areas in b) and c) indicate the timing of the ECH pulses from a). Dashed black lines in a) and b) mark the ECH pulse shown in c).

Note that since the timing of the ECH/ECCD pulses was performed uniformly for all discharges (Fig. 8a), the observed differences are not due to offset in the pulse timing. From Table I, we see that the RMP coil current, the ECH power, and the location of the ECCD current drive for shots 172325 and 172326 have similar values to other cases. The only parameter that differs considerably across the nine examined discharges is the ECCD current, which varies in the interval $I_{ECCD} = (45 - 409)\ kA$ (see Table I). However, the relationship between $T_{01}/T_{02}$ and $I_{ECCD}$ does not seem to follow a simple (linear or power law) trend. Instead, the transitional and reversed behavior of shots 172325 and 172326 seems to occur at intermediate values of the ECCD current $I_{ECCD} \approx (70 - 90)\ kA$. Thus, it is not clear if the observed phenomenon is caused by the value of $I_{ECCD}$. In the next section we compare data from experiments to vacuum field simulations, which suggest that the observed phenomenon is related to the characteristics of the island structure in these discharges.

## III. NUMERICAL ANALYSIS

A major goal of this study is to investigate how changes in the magnetic field topology affect the resulting cross-field electron transport. Here we consider three mechanisms affecting magnetic field topology: (i) shrinking of the large island around the $q = 1$ surface during the ECH/ECCD pulses, (ii) formation of stochastic region in the edge plasma due to island overlap, (iii) opening of additional small islands throughout the plasma. Due to the Hamiltonian nature of the magnetic field lines [47], these three mechanisms are not fully independent of each other. For example, opening of additional small islands can lead to island overlap in the edge plasma, which in turn produces an edge stochastic region. Vacuum simulations (discussed below) indicate that the transition to stochasticity leads to a shift in the location of large island chains throughout the plasma. For a fixed location of the ECH/ECCD pulse, the shifted location of the large island can change the efficiency of the island shrinking effect. Additional mechanisms, not considered in the present study, include dynamical processes, such as island bifurcation and island rotation. Those mechanisms are of high interest and will be explored in a future experimental campaign.

As the plasma response in the present experiments is assumed to be weak, we expect that conclusions about electron transport can be drawn from vacuum field simulations performed with the TRIP3D code [12] and the SURFMN code [48]. In TRIP3D, the positions of 8550 B-field lines in toroidal geometry are evolved under conditions informed from the experiments. The results include Poincaré plots of each discharge at a specific time of interest and histograms of the field line displacements as a function of their initial positions. In addition, SURFMN simulations were conducted to determine the number and width of magnetic islands for increasing mode number of the RMP perturbation and the vacuum island overlap width. Below we show representative plots illustrating key features of the magnetic field. Full set of experimental and simulation plots will be provided by the authors upon reasonable request.

### A. Large Islands Location With Respect to ECH/ECCD Pulse

As we saw in Sec. II, Fig. 8, in most examined cases, the ECH/ECCD pulses tend to enhance the suprathermal effect, which is visible by the time trace of the $T_{01}/T_{02}$ ratio. However, in shot 172325, this effect is diminished, and in shot 172326, the ECH/ECCD pulse leads to a drop of the $T_{01}/T_{02}$ ratio. One possibility is that the compared discharges had substantially different plasma conditions, perturbation parameters, or features of the ECH/ECCD pulses. All discharges were low density, inner-wall limited, with only Ohmic heating and no neutral beam heating. The possible role of changing plasma density in the core during ECH/ECCD pulses will be discussed in Sec. IV. Examination of Table I reveals that the 3D perturbation current and properties of the ECH/ECCD pulses used in shots 172325 and 172326 are very similar to those used in other shots. Specifically, the perturbation current increases from $0\ kA$ to $5.7\ kA$, the ECH power is about $2.2\ MW$ and the ECCD current is in the range $(70-90)\ kA$ is deposited at radial location $\rho_{ECCD} \in (0.21, 0.24)$. One notable difference is that shots 172325 and 172326 have the highest plasma current $I_p$ and the lowest $q_{95}$ values, which suggests that the radial location of the initial island structure (before RMPs are turned on) in these cases differ. This observation is interesting, as it suggests that, for a fixed location of the ECH/ECCD pulse, the location of the magnetic island chains can be manipulated to either enhance or diminish radial electron transport. Similarly, for a fixed distribution of island chains, we expect that differences in the electron transport will be observed when the tagging is performed at different locations with respect to the large islands. The latter type of experiment will be performed as part of a future experimental campaign.

In the present experiments, the ECH/ECCD deposition locations were similar across all shots, as seen from Table I. While in principle, DIII-D has the flexibility of independently varying $q_{95}$ for more-or-less fixed $q$-values in the core, in the present study, shots with large $q_{95}$ values have the $q = 1$ surface closer to the magnetic axis. In addition, these shots have steeper rate of change $q'$, which is why they can support larger number of rational surfaces, and therefore, more islands in the edge plasma. In contast, here, shots with smaller $q_{95}$ tend to support fewer larger island chains further away from the magnetic axis. Thus, since the location of the ECH/ECCD pulse is roughly fixed to $\rho \approx 0.1-0.2$ for all cases, it is expected that the electron tagging occurred at different locations with respect to the $q = 1$ island for shots with different $q_{95}$ values. In the presence of large magnetic islands, the radial electron diffusion should be slower than diffusion across unperturbed nested surfaces since the electrons need to diffuse across additional surfaces in the island interior. In addition, it has been experimentally demonstrated that electron heat diffusivity is reduced inside an island's O-point [49]. Thus, we expect that radial electron transport across islands can lead to electron sub-diffusion due to trapping effects. However, the motion of electrons trapped inside the separatrix of an island becomes ergodic and eventually all particles adopt the constants-of-motion of the exact resonance creating the island perturbation [1]. As a result, the energy of electrons trapped in surfaces across the island width does not drop with radial location. This effect should be more pronounced for wider islands.

This is evident from the flat regions in the ECE plots (Fig.4), which are assumed to coincide with island locations (e.g., consider the region around the $q = 2$ surface). While the ECE temperature stays constant throughout the width of the island, the temperature calculated from Thomson Scattering (TS) (green dashed line on Fig.4) drops across the same region. Since the TS and ECE have different poloidal and toroidal locations, it is possible that ECE is probing the island chains' O-points, while TS is probing X-points. It is also possible that the TS prediction yields dropping electron temperature because it is assuming Maxwellian electron distribution, which smoothens irregularities in the raw data. In either case, we expect that if the flat regions in ECE measurements coincide with island chain locations, when electrons escape the exterior surfaces of these islands, they will have higher energy than the energy they would normally have at the same radial location if the island perturbation was not there. In other words, islands can trap electrons, slowing down their cross-field diffusion, but electrons de-confined from islands will exhibit suprathermal transport in comparison to electrons that were not trapped by islands. The ECH/ECCD pulses used in the present experiments are a possible mechanism for such electron release from island surfaces.

ECCD techniques have been previously used to shrink magnetic islands [39], [40] if the current drive location is carefully chosen to coincide with the O-point of the island. If the current is driven off-axis, but inside an island, it is expected to still cause some shrinking due to perturbation of the separatrix surface and nearby surfaces. Based on the above logic, if the current drive location $\rho_{ECCD} < \rho_{q=1}$, where $\rho_{q=1}$ is the location of the $q = 1$ island axis, the tagged electrons will have to diffuse across the large $q = 1$ island, which should cause trapping and reduced suprathermal feature. In contrast, if $\rho_{ECCD} \gtrsim \rho_{q=1}$, the electrons can only be trapped in the smaller islands distributed at radius $\rho > \rho_{ECCD}$, and the suprathermal feature may not be appreciably affected. While the ECH/ECCD pulses are fairly localized, both the power deposition and the current drive have a finite spatial spread (see Fig. 3). To determine the spatial spread of electron tagging with respect to the spatial spread of the $q = 1$ island chain, TORAY simulations were used to determine the 3D location (in the radial, poloidal, and toroidal directions) of each pulse, while TRIP3D Poincaré maps were computed to identify the underlying vacuum field structure for each discharge. While it was determined that none of the pulse locations coincided exactly with the O-point of the $q = 1$ island, differences in suprathermal electron transport were observed when the tagging occurred at different locations with respect to $\rho_{q=1}$.

Figure 9 shows the TRIP3D Poincaré plots for several shots where the observed electron transport behavior differed during ECH/ECCD pulses. In shots 172322 and 172330 (Fig. 9a, b and red/magenta lines on Fig. 8), the suprathermal feature was enhanced during each pulse, while in shot 172326 (Fig. 9d and orange line on Fig. 8) the suprathermal feature was suppressed during each pulse. For shot 172325 (Fig. 9c and light green line on Fig. 8), the suprathermal feature was initially diminished, and subsequently slightly increased during the pulse, which is what we consider an intermediate behavior. Red rectangles on each plot in Fig. 9 indicate the location of the ECH/ECCD pulse with respect to a large magnetic island residing on the $q = 1$ surface (marked by a dashed red line). In the enhanced-transport cases, the location of the electron tagging is at $\rho_{ECCD} > \rho_{q=1}$ (Fig. 9a) or close to the X-point of the $q = 1$ island (Fig. 9b) and the radial "line of sight" (marked by extended light pink rectangles) passes through the X-points of other islands. In contrast, in the reduced-transport cases (Fig. 9c, d), the tagging occurs at $\rho_{ECCD} < \rho_{q=1}$ and the line of sight for the tagged electrons passes through thicker regions of neighboring islands. Examination of the other discharges confirms the trends observed in Fig. 9.

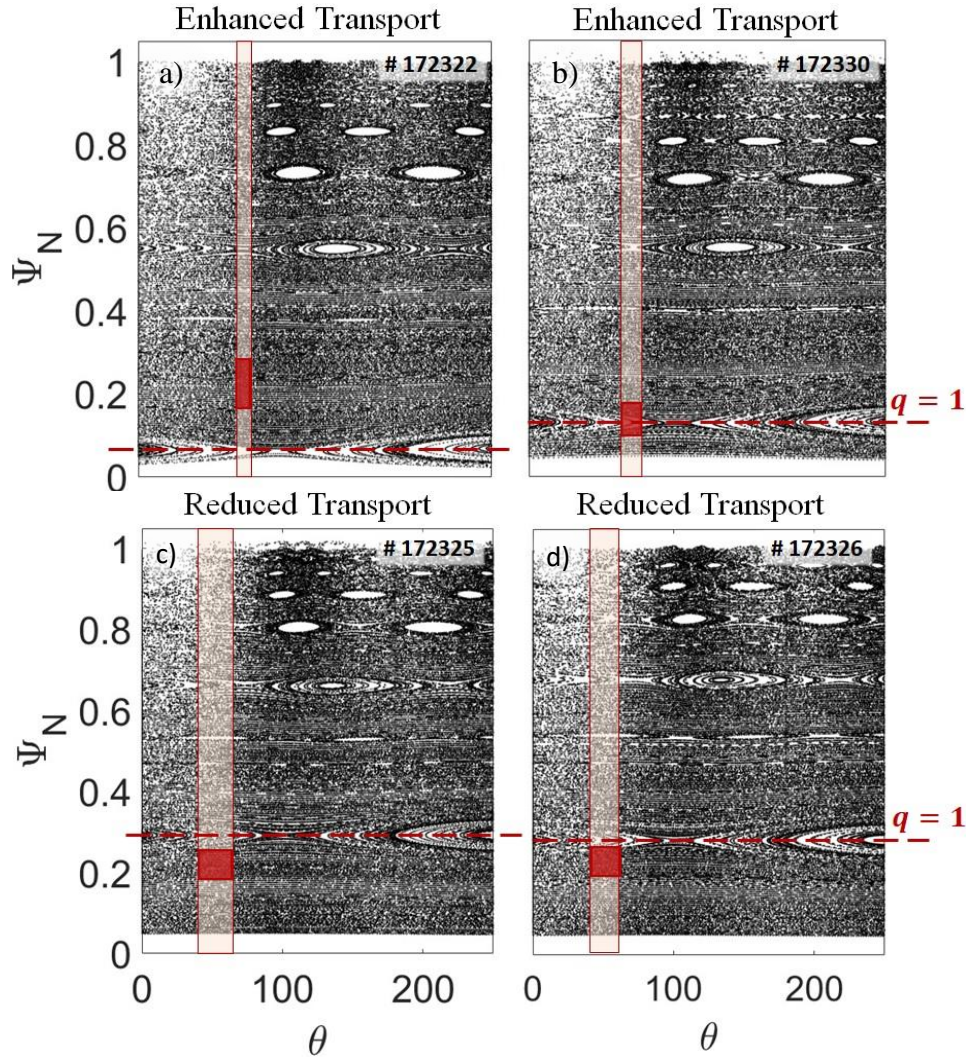

Fig. 9. TRIP3D Poincare plots of shot numbers a) 172322 and b) 172330, where enhancement of the suprathermal electron feature was observed during the ECH/ECCD pulse and shot numbers c) 172325 and d) 172326, where reduced transport was observed. All plots are generated for toroidal angle $\phi = 279°$ (TRIP3D coordinate system), which coincides with the location of the ECE diagnostic (81° in DIII-D machine coordinates).

These observations are compared against findings from Evans, et al. [50], where measurements of changes in the time it takes for modulated electron cyclotron heat pulses to propagate across the O-point of a 2/1 magnetic island in DIII-D were consistent with a model in which the island spontaneously bifurcates from smooth flux surfaces to a partially stochastic region. In other words, the heat transport was observed to be slower across the unperturbed island surfaces and faster across partially stochasicized regions. In the present study, we find evidence that suprathermal electron transport is suppressed when the electron cyclotron heat pulse diffuses across the large $q = 1$ island and enhanced when the pulse diffuses across island X-points, where local stochasticity is expected to occur. Although these results are preliminary, they hint towards the possible role of magnetic islands into acceleration and trapping of electrons. This will be further explored by the authors in a next experimental campaign where the ECH/ECCD pulse will be delivered at different locations with respect to the large island chains.

## B. Transition to Stochasticity

It has been proposed that raising RMP coil currents can induce a transition from a regime dominated by nested magnetic flux surfaces through a regime with a variety of magnetic islands to a regime where the islands have "overlapped" to produce an extended region of stochastic field lines. A regime with ideal nested flux surfaces is not accessible in the present experiments even at $I_{RMP} = 0 \, kA$ as magnetic perturbations from intrinsic error fields and EFC coils always lead to the formation of island structures. In particular, even in the absence of RMPs, all examined experiments feature at least two large island chains, one at the $q = 1$ surface and another one at the $q = 2$ surface. To produce a region of stochastic field lines in the edge plasma, RMPs were used to create and grow islands at the $n = 3$ surfaces until island overlap is achieved. We observe that increasing the RMP coil current leads to both the formation of new smaller islands and stochastization of the large islands' exteriors. As discussed below, the critical current value for the transition to stochasticity is dependent on the $q_{95}$ for each discharge.

Possible uncertainties in the calculation of the magnetic islands result from the uncertainties in the location of the resonant surfaces (uncertainties in safety factor profile in Grad-Shafranov equilibrium reconstruction) and the uncertainties in the estimation of the magnitude of the perturbation field in the plasma (uncertainties in the measurements of coil currents and representation of the coil geometries as simplified thin wires). Given the stochastic nature of the fields near the island chains and in the plasma edge, detailed quantification of uncertainties is quite complicated and lies outside of the scope of this paper.

Figure 10 shows histograms of the B-line displacements from their initial position computed with TRIP3D using initial conditions from shot 172322. In this discharge, the $I_{RMP}$ was varied from $0 \, kA$ to $5 \, kA$ in three steps (red line on Fig. 2c). The colorbar in each plot from Fig. 10 represents a log scale of the normalized number of times a B-line intersected a given location in space. In this representation, narrow red line regions correspond to nested magnetic surfaces, or locations in space where the lines did not deviate signficantly from their initial positions as the simulation was advanced. Wide rhombus regions on the plots indicate the locations of magnetic island chains, and irregular blue regions suggest stochastization of the field lines.

In Fig. 10a, for $I_{RMP} = 0 \, kA$, the B-line histogram shows two big islands, located at the $q = 1$ and $q = 2$ surfaces, and $4 - 5$ smaller, but distinct islands located between the $q = 2$ surface and the edge plasma. These initial islands are created by the intrinsic error fields in DIII-D (shifted and tilted pooloidal fied coils and B-field buswork) and by the $n = 1$ error field correction currents (in the DIII-D C-coils). The values for the intrinsic fields in DIII-D have been previously masured and discussed in more detail in [27], [31], [51]. As the current in the RMP-coils is increased to $I_{RMP} = (2 - 3) \, kA$ (Fig. 10b, c), new island chains emerge between the $q = 1$ and $q = 2$ surfaces and the previous island chains develop slightly stochastic exterior (blue color around rhombus structures). For $I_{RMP} = 5\ kA$ (Fig. 10d), more island chains emerge in the edge region and stochastic exteriors are enhanced. However, even at $I_{RMP} = 5\ kA$, the island chains in shot 172322 remain distinct all the way to the edge and no significant formation of a connected stochastic layer is observed.

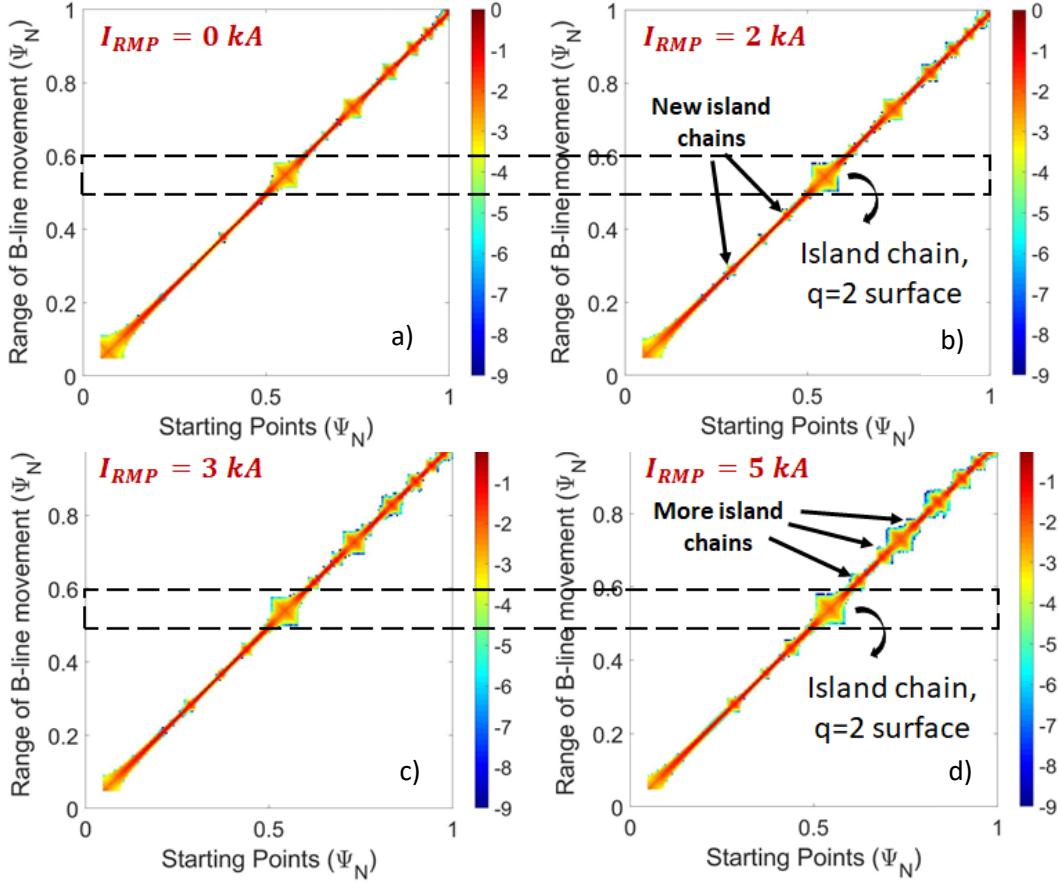

Fig. 10. Shot 172322: histogram of B-line displacements from the original position a) at $2600\ ms$ when $I_{RMP} = 0\ kA$, b) at $3100\ ms$ when $I_{RMP} = 2\ kA$, c) at $4100\ ms$ when $I_{RMP} = 4\ kA$, and d) at $5100\ ms$ when $I_{RMP} = 5\ kA$. The color bar is a log scale of the normalized number of field lines, crossing a given location in space.

The transition to a regime where the edge islands overlap, resulting in a region of stochastic field lines, is visible when higher RMP current is applied. Figure 11 shows that at $I_{RMP} = -6\ kA$ for shot 172330 (Fig. 11a) and at $I_{RMP} = 5.7\ kA$ for shot 172326 (Fig. 11b), a stochastic layer forms in the region $\Psi_N = (0.8 - 1)$. These regions are not fully stochastic as remnants of island core structures are still visible in both cases. However, the stochastic exterior of these former island chains now surrounds all of them in the edge region. These results are further confirmed from SURFMN simulations, which show that an overlap of the edge island chains occurs at $\Psi_N = 0.84$ for $I_{RMP} = -6\ kA$ in shot 172330 (Fig. 11c) and at $\Psi_N = 0.80$ for $I_{RMP} = 5.7\ kA$ in shot 172326 (Fig. 11d). Comparison of TRIP3D histograms and SURFMN plots for the rest of the examined experiments reveals that a stochastic edge region is also observed to appear in shots 172324-172328 at $I_{RMP} = 5.7\ kA$ and in shot 172329 at $I_{RMP} = -6\ kA$, suggesting that the phenomenon is repeatable and observable with both positive and negative currents.

For the $n = 3$ perturbations used here, the six I-coils are energized in $+ - + - + -$ or $- + - + - +$ patterns, and what is shown here for $I_{RMP}$ is the value in the first of these coils (the one located at 30 degrees toroidally in DIII-D). Thus, the difference between positive and negative $I_{RMP}$ is shifting the perturbation by 60-degree toroidally, which results in shifted toroidal locations of the island O-points and X-points, but no difference in the number, width, or radial location of the island chains. The phase of the $I_{RMP}$ (0 or 60 degrees) changes how that perturbation couples to the intrinsic and EFC fields (which are fixed toroidally).

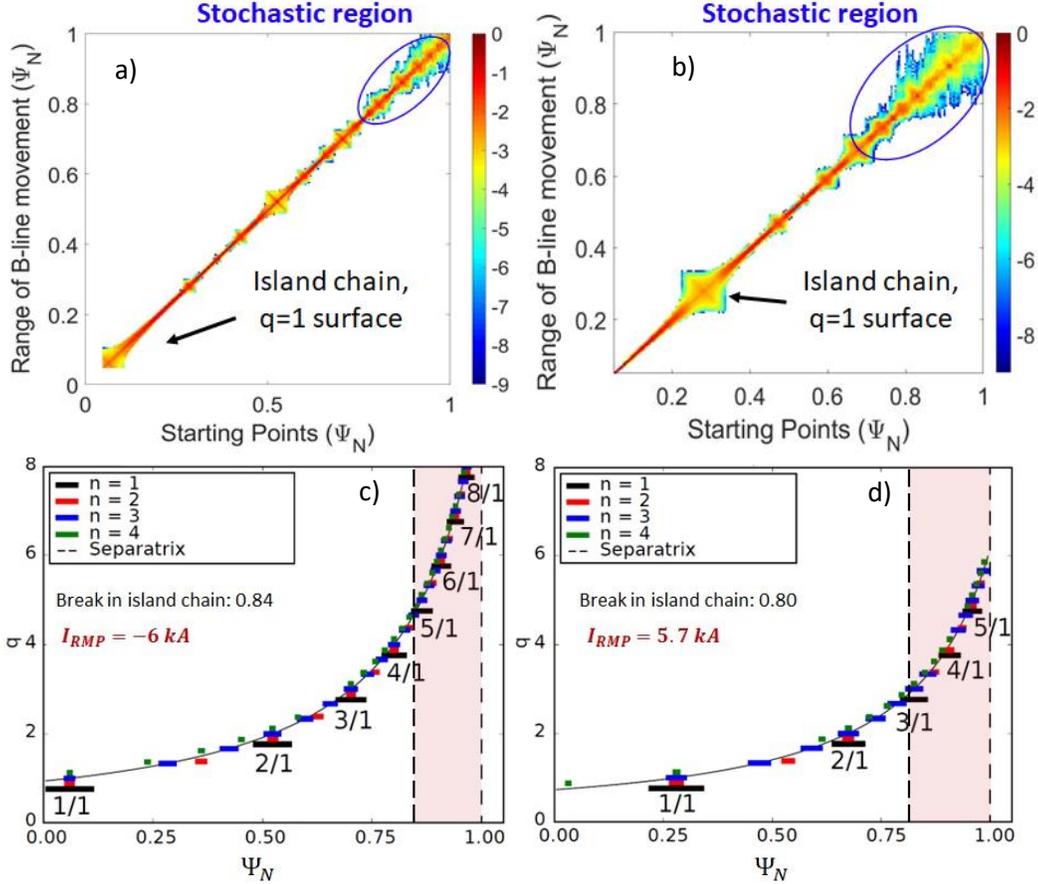

Fig. 11. TRIP3D histograms of B-line spread at $3100\ ms$ for a) shot 172330 and b) shot 172326. SURFMN plots of island width at $3100\ ms$ for c) shot 172330 and d) shot 172326 Shaded regions in c) and d) indicate the width of stochastic region, where island surfaces overlap.

Examination of similar histograms calculated for the rest of the experiments suggests that the RMP current threshold needed to transition from individual islands with small stochastic exteriors (Fig. 10d) to island remnants surrounded by common stochastic region (Fig. 11a, b) varies with the q-profile for each shot. The present experiments were not performed with small enough $I_{RMP}$ steps to pinpoint the exact threshold value for each q-profile. However, for a fixed type of perturbation (here $n = 3$ perturbation from RMP coil), shots with smaller $q_{95}$ values are observed to form wider stochastic regions in the edge plasma. This makes sense as the smaller $q_{95}$ means that fewer but larger islands are present close to the edge and small island growth can lead to overlap. However, cases with higher $q_{95}$ value can support more perturbation modes in the edge, i.e., more island chains. In these cases, a transition to stochasticity can be achieved by a combination of perturbations from different coils, which can be desirable as it reduces the threshold current requirement for each coil. Finally, we note that once the stochastic regions form (Fig. 11a, b), the locations of the $q = 1$ and $q = 2$ islands shift towards smaller $\Psi_N$. This suggests that if the observed electron transport is dependent on the location of $q = 1$ island chains with respect to the ECH/ECCD pulses (as discussed in Sec. III A), the controlled formation of a stochastic region in the edge plasma can be used as a method to manipulate electron transport.

## IV. DISCUSSION

### A. The Role of Changing Plasma Density

In Sec. II we distinguished two types of energetic electrons: relativistic runaway electrons that occur during the first ECH/ECCD pulse for most discharges, and electrons that are neither thermal nor relativistic, observed during subsequent ECH/ECCD pulses. The former resulted in both enhanced ECE temperatures and signals on the HXR detectors, which suggest that their energy is $\gtrsim 5\ MeV$. The latter type of energetic electrons caused elevated ECE temperature profiles with suprathermal feature in the edge plasma measurements, but they did not yield substantial peaks on the HXR detectors, suggesting that the energy of these electrons is $< 5\ MeV$. We quantified the nonthermal feature using the ratio $T_{01}/T_{02}$, where $T_{01}(T_{02})$ is the temperature of ECE chord 1 (chord 2). A value $T_{01}/T_{02} > 1$ indicates the presence of suprathermal electrons and the bigger the ratio, the more pronounced is the nonthermal feature. In Fig. 8, we showed that the ECH/ECCD pulses enhance the ratio $T_{01}/T_{02}$ for most discharges but decrease the ratio for shots 172325 and 172326. In this section, we discuss the possible role of changing plasma density in the observed transport differences.

DIII-D experiments using the Gamma Ray Imager (GRI) System [44], [45] have suggested that the energetic electron distribution $f_e$ for low-density (Ohmic heating only) discharges exhibit a non-monotonic feature as a function of electron energy $E_e$, with a peak location dependent on the plasma density (see Fig. 11 from [44]). The energies recorded in those experiments were in the range $E_e \in (3 - 16)\ MeV$. That study showed that as the plasma density was increased from $n_e = 1 \times 10^{19}\ m^{-3}$ to $n_e = 1.5 \times 10^{19}\ m^{-3}$, the whole $f_e$ profile shifted downwards (i.e., less counts at all energies) and the distribution peak location shifted from $7\ MeV$ to $5\ MeV$. In other words, for such discharge conditions an increased plasma density leads to suppression of the suprathermal features by both decreasing the total number of energetic electrons detected and the peak $E_e$ value. In the present experiments, where the plasma conditions are similar to those in [44], the electron density $n_e$ was measured using Multichannel CO2 Interferometer (MCI) and Thomson Scattering (TS) data [38]. In the absence of ECH/ECCD pulses, the mid-plane, line-averaged plasma density obtained from MCI data [52] (Fig. 12a) shows electron densities within the range $n_e \approx (1 - 1.5) \times 10^{13} cm^{-3}$, which is similar to the densities examined in [44]. Similar density range was observed from TS data.

Comparison to scintillator data (Fig. 7) suggests that the shots with lowest electron density (1672327-172330) exhibit the most pronounced peaks, i.e., highest count of relativistic runaway electrons, during the first ECH/ECCD pulse. This observation is in agreement with the findings from [44]. However, in Fig. 7 we also observed scintillator data peaks prior to the first ECH/ECCD pulse for discharges 172325 and 172326, which have higher line-averaged electron density. Therefore, we conclude that plasma density alone cannot account for the observation of runaway electrons in our study. We further observe that during all ECH/ECCD pulses, the electron density increases, on average, by $0.5 \times 10^{13} cm^{-3}$. The increased density during ECH appears to result partially from increased carbon influx due to higher power flux to the walls based on partially ionized edge carbon emission and core Zeff measurements. The suprathermal features are enhanced for most cases, except for discharges 1723225 and 172326. For these two cases, line-averaged MCI densities measured in the edge plasma suggest that shots 172325 and 172326 consistently had the highest electron density in the edge, which is expected as $n_e \sim I_p$, all else being equal. This may be one reason why for these shots the suprathermal feature in Fig. 8 was suppressed during ECH/ECCD pulses. However, high density in the edge does not explain the observation of runaway electron peaks and suprathermal tails in the absence of ECH/ECCD for these two shots. Thus, we conclude that, while electron density is an important factor, it does not explain all features of the nonthermal electron transport observed here.

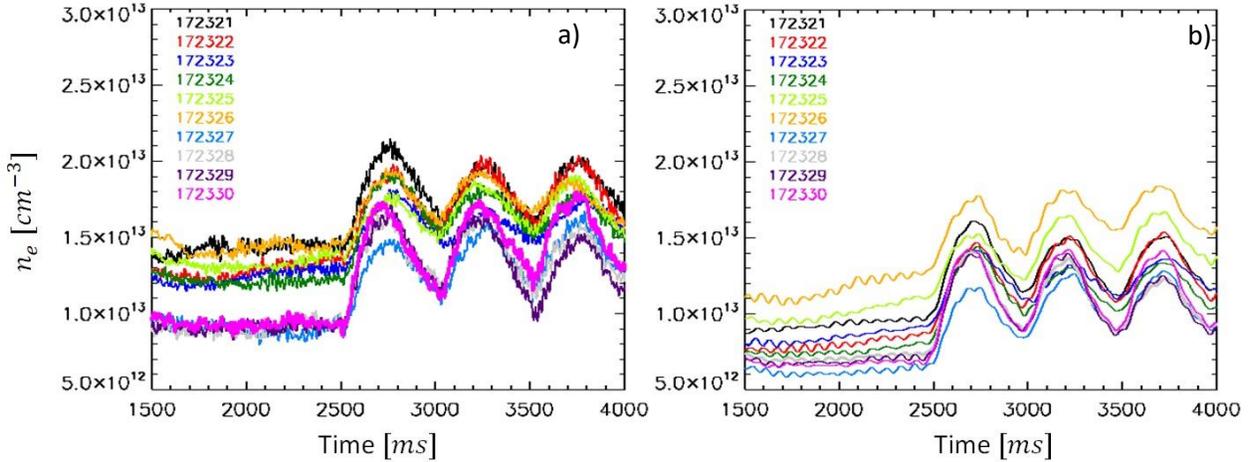

Fig. 12. Electron density from MCI a) mid-plane, line-averaged data b) line-averaged data from a chord viewing the edge plasma.

B. The Role of RMP Perturbations

The vacuum field simulation results from Sec. III suggest that suprathermal electron transport is enhanced or suppressed based on the deposition location of the ECH/ECCD pulse with respect to the $q = 1$ island chain. Specifically, for similar radial locations of the ECH/ECCD pulse, comparison of ECE signals to vacuum field simulations suggests that suprathermal transport is diminished when $\rho_{ECCD} < \rho_{q=1}$ (shots 172325 and 172326), possibly due to electron trapping in the $q = 1$ island chain. We further discussed that the formation of a stochastic region in the edge plasma due to sufficient RMP perturbations can lead to a shift in the location of the unperturbed island chains in the core plasma. Thus, the RMPs can affect electron transport in two ways: (i) directly – by eliminating small islands in the edge plasma (reducing small trapping effects and enabling chaotic trajectories) and (ii) indirectly – by moving the core islands to smaller radial locations (reducing large trapping effects).

Since both mechanisms are expected to reduce the island trapping effects, we expect that increasing the width of the stochastic region should lead to increased suprathermal effect. Figure 13 shows a scatter plot of the ratio $T_{01}/T_{02}$ as a function of the vacuum islands overlap width (*viow*) [53] for discharges that simultaneously had nonzero RMP and the ECE chords 01 and 02 located in the edge plasma. While not ideal, an almost linear relationship can be observed at small time scales of $\approx 20\ ms$, where an increasing *viow* leads to an enhanced $T_{01}/T_{02}$ ratio, supporting the hypothesis that wider stochastic region in the edge can lead to enhanced electron transport. Note, however, that the width of the stochastic region does not only depend on the current amplitude in the RMP coils. For a fixed current amplitude, *viow* further depends on the initial field topology, which is discussed later in this section.

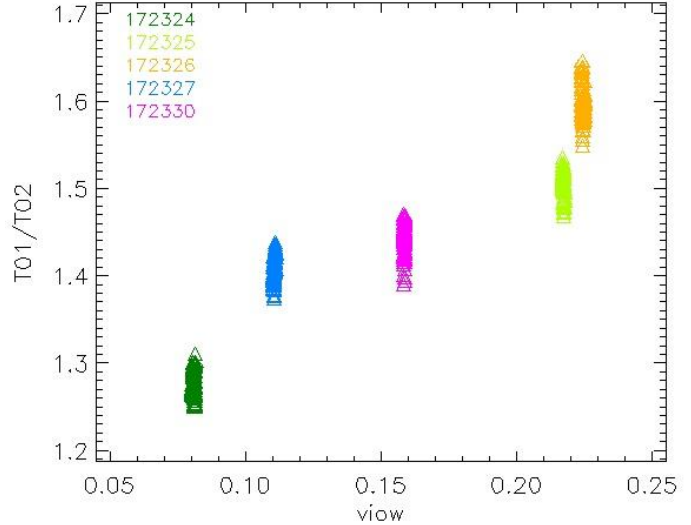

Fig. 13. Vacuum islands overlap width (viow) in units of normalized poloidal flux versus the ratio $T_{01}/T_{02}$ during RMP coil perturbation. $I_{RMP} = 5.7 kA$ for shots 172324-172327 and $I_{RMP} = -6kA$ for shot 172330. Data here was collected during the time interval

Deviations from a linear trend in Fig. 13 result from several inconsistencies among the discharges. First, the ECH power and the ECCD current drive are not identical for all shots. Specifically, the typical ECH power for pulses in shot 172330 (magenta color on Fig. 8a) was $2.7 - 3\ MW$, while the power in shot 172324 was as low as $1.6\ MW$ and about $2.2\ MW$ for other shots. Additionally, as seen from Fig. 8a, the pulses in shot 172330 exhibit large fluctuations (~$10\ ms$ drops in power), which result in fluctuations in the $T_{01}/T_{02}$ since efficiency of the electron tagging is proportional to ECH power. This leads to a spread of $T_{01}/T_{02}$ values for timescales longer than $\gtrsim 20\ ms$. We further observed that over a time interval of $\approx 100\ ms$, the value of *viow* for shots 172325 and 172326 can collapse to zero and subsequently increase back to about $0.2 - 0.25$ width without substantial change in the ratio $T_{01}/T_{02}$. In contrast, the width of the stochastic region for shots 172324, 172327, and 172330 remain roughly constant over the same time interval.

One possible explanation is that for shots 172325 and 172326, the $q = 1$ island collapses and reopens or bifurcates. During the island collapse, the radial location of smaller island chains is shifted away from the edge plasma and towards the core plasma. This, in turn, decreases the island overlap and closes the stochastic region in the edge. Alternatively, on a timescale of $\approx 100\ ms$, the island may bifurcate from smooth flux surfaces to a partially stochastic region due to a nonlinear amplification of secondary resonant islands, which has been observed for ECH pulses close to a $q = 2$ island in DIII-D [50]. The collapse of the stochastic region and its relation to island bifurcation or collapse are topics of great interest and will be examined in subsequent experimental campaign. Despite these differences, the trend in Fig. 13 provides initial evidence that wider stochastic region in the edge plasma results in enhanced nonthermal electron transport.

Data on Fig. 13 corresponds to time periods where the magnitude of the RMP coil currents was similar for the examined discharges. The RMP current in shot 172330 was $I_{RMP} = -6\ kA$, while the other shots had $I_{RMP} = 5.7\ kA$. As can be seen on fig. 13, similar $I_{RMP}$ leads to different widths of the edge stochastic region, which is likely due to the differences in the $q_{95}$ profile. Fig. 14a shows a plot of *viow* as a function of $q_{95}$ for a time interval without RMP (unfilled diamonds) and a time interval with RMP (filled diamonds). As can be seen, in the absence of RMP, the discharges with the largest $q_{95}$ have the largest *viow*. This makes sense since discharges with large $q_{95}$ already support more islands in the edge (due to other coil perturbations), which leads to higher probability for overlap even without the additional RMP. When RMPs with similar $I_{RMP}$ are applied, the discharges with the smallest $q_{95}$ form the largest *viow*. Without the perturbation, those discharges support fewer islands in the edge plasma, but those islands are larger. Thus, when the RMPs are applied, a modest growth of the larger islands can lead to closing of the gaps between them and trigger the onset of stochasticity due to island overlap.

Figure 14b shows the plot of $T_{01}/T_{02}$ as a function of $q_{95}$ for a time interval with perturbation coils on and similar RMP currents. This plot supports the observation that, for similar $I_{RMP}$, discharges with smaller $q_{95}$ exhibit more pronounced suprathermal feature, possibly through the formation of a wider edge stochastic region. The relationship between safety factor and edge plasma transport has been previously discussed in the context of ELM suppression studies in DIII-D. ELMs, or edge localized modes, are potentially damaging pedestal instabilities in the edge tokamak plasmas that can be caused by edge current density and/or pressure driven peeling-ballooning instabilities [54]. It has been demonstrated [3] that ELM suppression in H-mode plasma can be correlated with a threshold width of the edge stochastic region. Similar to the present work, the stochastic region in [3] is defined as the edge region having magnetic islands with Chirikov parameter $> 1$ (i.e., island overlap), based on vacuum calculations excluding the self-consistent plasma response. The threshold width of the stochastic region needed for the ELM suppression in [3] was found to vary with the safety factor profile. Depending on the $q_{95}$, the threshold width could be achieved by increasing the $n = 3$ RMP field strength or by varying the combinations of $n = 3$ and $n = 1$ perturbations.

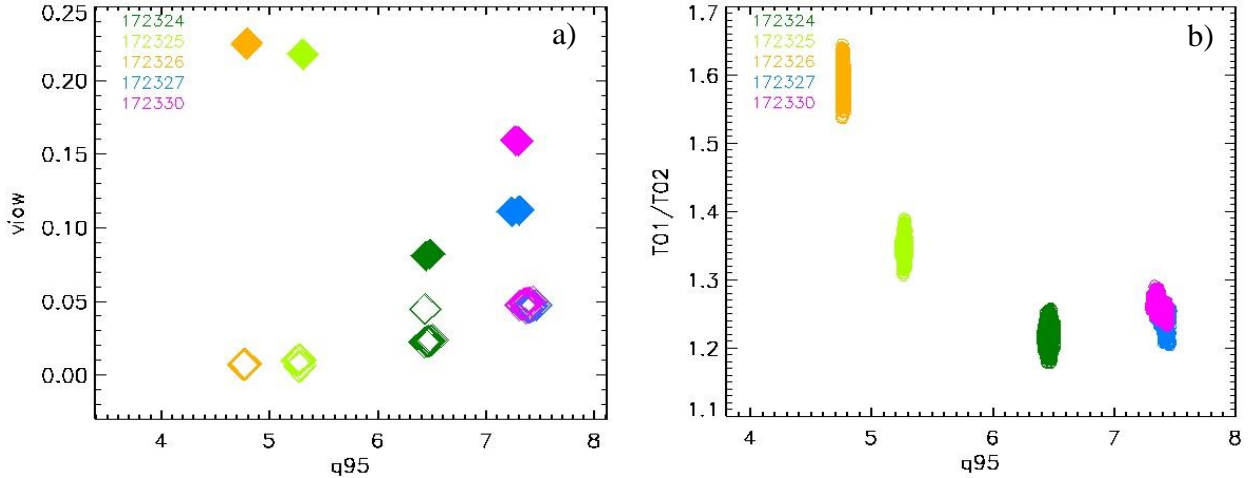

Fig. 14. a) Vacuum islands overlap width (viow) versus $q_{95}$. Unfilled symbols correspond to time with no RMP, while filled symbols correspond to times when the RMP coils are on. b) the ratio $T_{01}/T_{02}$ versus $q_{95}$ during RMP coil perturbation. For both plots, when RMP is on, $I_{RMP} = 5.7 kA$ for shots 172324-172327 and $I_{RMP} = -6 kA$ for shot 172330.

While the present experiments are using low confinement (L-mode) discharges and the studied transport is suprathermal electrons (versus ELMs in H-mode plasmas in [3]), future extension of the present study may be helpful in explaining the physical mechanisms guiding the correlation between ELMs suppression and width of the edge stochastic region. The study in [3] demonstrates that increasing width of the edge stochastic region leads to ELM suppression, while in the present study, we find that increasing *viow* leads to enhanced radial transport of suprathermal electrons. Thus, an initial hypothesis can be formed suggesting that release of suprathermal electron populations through radial transport may result in mitigation of edge/pedestal instabilities, such as ELMs. The possible relationship between energetic electrons and ELMs is not a new idea. For example, measurements of microwave and x-ray emission during ELM activity in the MAST tokamak indicate the presence of suprathermal, magnetic-field-aligned electron populations [55]. Therefore, an interesting question for future research is understanding the complex relationships between energetic electrons, ELM activity, and edge plasma stochasticity. As electron kinetic effects are typically not incorporated in fluid models for the instability that drives ELMs, data from L-mode experiments, such as the those presented here, can prove valuable.

## V. CONCLUSIONS AND NEXT STEPS

Here we presented the results from DIII-D experiments, where energetic electron transport was investigated using an electron tagging technique. Each considered shot was an inner-wall limited L-mode discharge, using only Ohmic heating (no neutral beam injection). The plasma densities were intentionally kept low to avoid nonlinear plasma response and to allow for comparison between diagnostic measurements and vacuum field simulations. ECH/ECCD pulses were used to cause resonance with electrons within narrow regions in the core plasma. Then, the spatial and temporal spread of the tagged electrons was observed on ECE and HXR diagnostics. Two types of energetic electrons were identified: relativistic (runaway) electrons on energy $\gtrsim 5\ MeV$ and electrons that were not relativistic, yet, non-thermal with energies $< 5\ MeV$. To improve the accuracy of energy measurements, the team plans to conduct future experiments, where additional diagnostics are employed, including gamma ray imaging (GRI), improved scintillators (HXR), and visible synchrotron emission (SE). We further plan to analyze the present data using a fractional Laplacian spectral approach [56], [57] to better understand the interplay between nonlocality and stochasticity in the field topology and the resulting anomalous electron diffusion. Nevertheless, the present study is an important first step that demonstrates that the combination of electron tagging, diagnostic measurements, and vacuum field simulations can be used to probe the local transport of electron populations located within a narrow region within the plasma discharge. Specifically, of future interest will be probing electron transport at locations where interesting magnetic field topology (islands/stochastization) are expected to occur.

Another interesting result is the observed suppression of the electron suprathermal feature for discharges where the electron tagging occurred at radial locations smaller than the location of the large $q = 1$ island chain. We attribute this reduced transport to trapping effects caused by the island chain. A future experimental campaign led by the authors will aim to quantify the relationships between electron transport and island topology, including island width, location, structure (e.g., X-points vs. O-points), and dynamics (island shrinking or bifurcation). This future study will aim to reproduce similar experimental scenarios, but extra care will be taken to minimize the effects of plasma density fluctuations and spatial spread of the ECH pulses. Another observation is that the large island chains shift location when the RMP coil is on, which results from emergence of new island chains and the increased width of the stochastic region in the edge. Therefore, we conclude that the 3D coils can be used to manipulate the location of islands with respect to the location of ECH/ECCD pulses, which in turn is expected to enhance or suppress the suprathermal electron feature.

Finally, we established that the prominence of energetic electron features on the ECE data is proportional to the width of the RMP-induced edge stochastic region. We further observe that in the absence of RMP perturbation, discharges with larger $q_{95}$ are more likely to form an edge stochastic layer due to their ability to support many small islands in the edge, resulting from various coil perturbations. However, for fixed nonzero amplitude of the RMP coil current, we observe that discharges with smaller $q_{95}$ can form wider regions of edge stochastic fields due to easier overlap of large island chains located closer to to the edge. Thus, we expect that when edge stochastic region is desired, a combination of coil perturbations should be used for discharges with large $q_{95}$, while an increasing RMP current amplitude may be more effective for cases with smaller $q_{95}$. These findings may be useful for future experiments attempting to control energetic electron transport or the occurrence of ELMs using plasma edge stochasticity.

## ACKNOWLEDGEMENTS

This material is based upon work supported by the U.S. Department of Energy, Office of Science, Office of Fusion Energy Sciences, using the DIII-D National Fusion Facility, a DOE Office of Science user facility, under Awards DE-SC0023061, DE-FG02-05ER54809, DE-SC0021405, DE-FG02-97ER54415, and DE-FC02-04ER54698.

The authors thank S. H. Nogami, W. W. Heidbrink, and J. T. McClenaghan for assistance and useful discussions.